\documentclass[preprint,12pt]{elsarticle}
\usepackage{amsmath}
\usepackage{amssymb}

\input{tcilatex}

\begin{document}

\begin{frontmatter}



\title{
The reduction of qualitative games
}

\author{Monica Patriche}

\address{
University of Bucharest
Faculty of Mathematics and Computer Science
    
14 Academiei Street
   
 010014 Bucharest, 
Romania
    
monica.patriche@yahoo.com }

\begin{abstract}
We extend the study of the iterated elimination of strictly dominated strategies (IESDS) from
 Nash strategic games to a class of qualitative games. Also in this case, the IESDS process 
leads us to a kind of ´rationalizable´ result. We define several types of dominance relation 
and game reduction and establish conditions under which a unique and nonempty maximal 
reduction exists. We generalize, in this way, some results due to Dufwenberg and Stegeman 
(2002) and Apt (2007).
\end{abstract}


\end{frontmatter}



\label{}





\bibliographystyle{elsarticle-num}
\bibliography{<your-bib-database>}







\section{INTRODUCTION}

Bernheim [3] and Pearce [16] studied the rationalizable strategic behavior
in the framework of non-cooperative strategic games introduced by Nash [14].
The rational behavior of the players is a fundamental assumption in Game
theory. It implies that each strategic game can be characterized by a
process of iterated elimination of strictly dominated strategies (IESDS).
The result of this process is known as the maximal reduction of the game.

The iterated elimination of strictly dominated strategies has several
different definitions. We must refer to the approaches of Gilboa, Kalai and
Zemel [9,10], Milgrom and Roberts [13], Marx and Swinkels [12], Ritzberger
[17], Dufwenberg and Stegeman [8], Chen, Long and Luo [5], or Apt [1,2] as
some important ones in the literature. Osborne and Rubinstein [15] and
Rubinstein [18] also developed some topics concerning rationality.

The main problems concerning the IESDS procedure are related to the
non-emptiness and uniqueness of the limit game. In the case of infinite
games, the order of reductions is important, and the maximal reduction may
not be unique if different paths are considered. Dufwenberg and Stegeman [8]
proved the uniqueness and nonemptiness of the maximal reduction for a
strategic game with compact strategy sets and continuous payoff functions.
Apt [1] treated the various definitions of IESDS in a unitary way,
specifying the games where the definitions coincide. His approach is based
on complete lattice and the study of operators.

In order to develop the ideas concerning the rationality, we consider a
model which generalizes the strategic game. We consider the qualitative
games which have, for every player, a strategy set and a preference
correspondence constructed by using the utility functions. Nash's
equilibrium point is seen in this framework as a maximal element. We also
consider different types of majorized correspondences, which generalize the
well-known semicontinuous ones. So, we work with U-majorized correspondences
defined by Yuan and Tarafdar [20], $Q_{\theta }-$majorized correspondences
introduced by Liu and Cai [11] and L$_{S}$-majorized correspondences due to
G. X. Yuan [19]. We use theorems which prove the existence of maximal
elements for qualitative games having these types of considered
correspondences. These results are due to Ding [7], Liu and Cai [11], and
Chang [4].

Our new approach wants to emphasize that the IESDS process leads to a kind
of $%
{\acute{}}%
$rationalizable$%
{\acute{}}%
$ result in the extended games. By changing the context, we underline the
idea of rationality, obtained in an iterated process of elimination the
unfitted strategies. We want to highlight the concept, rather than the
context where it was initially defined. We introduce several types of
dominance relation and game reduction and establish conditions under which a
unique and nonempty maximal reduction exists. In this way, we generalize,
some results due to Dufwenberg and Stegeman [8] and Apt [1]. On the other
hand, the examples of Dufwenberg and Stegeman [8] involve games with
discontinuous utility functions and this fact gives us the idea that the
problem of order independence can be raised for qualitative games, which can
also be generalizations of discontinuous strategic games.

The paper is organized in the following way: Section 2 contains
preliminaries concerning topological properties of correspondences and
qualitative games. The main results are presented in Section 3, after the
subsections containing the introduction of definitions, the problem of game
reduction and order independence and the conditions under which order
independence is obtained. The last subsection is dedicated to proving that
the set of maximal elements is preserved in any game by the process of
iterated elimination of strictly dominated strategies. Concluding remarks
are presented at the end.

\section{PRELIMINARIES AND NOTATIONS}

\subsection{TOPOLOGICAL PROPERTIES OF CORRESPONDENCES}

Throughout this paper, we shall use the following notations and definitions:

Let $A$ be a subset of a topological space $X$. 2$^{A}$ denotes the family
of all subsets of $A$. cl $A$ denotes the closure of $A$ in $X$. If $A$ is a
subset of a vector space, co$A$ denotes the convex hull of $A$. If $F$, $T:$ 
$A\rightarrow 2^{X}$ are correspondences, then co$T$, cl $T$, $T\cap F$ $:$ $%
A\rightarrow 2^{X}$ are correspondences defined by $($co$T)(x)=$co$T(x)$, $($%
cl$T)(x)=$cl$T(x)$ and $(T\cap F)(x)=T(x)\cap F(x)$ for each $x\in A$,
respectively. The graph of $T:X\rightarrow 2^{Y}$ is the set Gr$%
(T)=\{(x,y)\in X\times Y\mid y\in T(x)\}\medskip $

We present the notion of a compactly open (compactly closed) set, introduced
by Ding [6].

\begin{definition}
(Ding, [6]) A nonempty subset $D$ of a topological space $X$ is said to be 
\textit{compactly open} (respectively, \textit{compactly closed}), if for
every nonempty compact subset $C$ of $X$, $D\cap C$ is open (respectively,
compactly closed) in $C$. The \textit{compact interior} of $D$ is defined by
cint$D$ =\{$G:G\subseteq D$ and $G$ is compactly open in $X$\}. It is easy
to see that cint$D$ is a compactly open set in $X$ and for each nonempty
compact subset $C$ of $X$ with $D\cap C=\emptyset $, we have $($cint$D)\cap
C=$int$C(D\cap C),$ where int$C(D\cap C)$ denotes the interior of $D\cap C$
in $C$. It is clear that a subset $D$ of $X$ is compactly open in $X$ if and
only if cint$D=D$.$\medskip $
\end{definition}

Several classical types of continuity for correspondences are given.$%
\medskip $

DEFINITION 1. Let $X$, $Y$ be topological spaces and $T:X\rightarrow 2^{Y}$
be a correspondence

\QTP{Body Math}
1. $T$ is said to be \textit{upper semicontinuous} if, for each $x\in X$ and
each open set $V$ in $Y$ with $T(x)\subset V$, there exists an open
neighborhood $U$ of $x$ in $X$ such that $T(y)\subset V$ for each $y\in U$.

2. $T$ is said to be \textit{lower semicontinuous} if, for each x$\in X$ and
each open set $V$ in $Y$ with $T(x)\cap V\neq \emptyset $, there exists an
open neighborhood $U$ of $x$ in $X$ such that $T(y)\cap V\neq \emptyset $
for each $y\in U$.

3. $T$ is said to have \textit{open lower sections} if $T^{-1}(y):=\{x\in
X:y\in T(x)\}$ is open in $X$ for each $y\in Y.$\medskip

\subsection{QUALITATIVE GAMES}

Let $I$ be a non-empty set (the set of agents). For each $i\in I$, let $%
G_{i} $ be a non-empty topological vector space representing the set of
actions, let $P_{i}:\tprod_{i\in I}G_{i}\rightarrow 2^{G_{i}}$ be the
preference correspondence and $u_{i}:\tprod_{i\in I}G_{i}\rightarrow \mathbb{%
R}$ be the utility function.

\textit{Notation.} For each $i\in I,$ let us denote $G_{-i}:=\tprod%
\nolimits_{j\in I\setminus \{i\}}G_{j}.$ If $x\in \tprod_{i\in I}G_{i},$ we
denote $x_{-i}=(x_{1},...,x_{i-1},x_{i+1},...)\in G_{-i}.$ If $x_{i}\in
G_{i} $ and $x_{-i}\in G_{-i}$, we shall use the notation $%
(x_{i},x_{-i})=(x_{1},...,x_{i-1},x_{i},x_{i+1},...)=x\in \tprod_{i\in
I}G_{i}.$

\begin{definition}
(Nash, [14]) The family $\Gamma =(G_{i},u_{i})_{i\in I}$ is said to be a 
\textit{strategic game.}
\end{definition}

\begin{definition}
(Nash, [14]) An \textit{equilibrium} for $\Gamma $ is defined as a point $%
x^{\ast }\in \tprod_{i\in I}G_{i}$ such that for each $i\in I$, $%
u_{i}(x^{\ast })\geq u_{i}(x_{i},x_{-i}^{\ast })$ for each $x_{i}\in G_{i}$%
.\medskip
\end{definition}

\begin{definition}
The family $G=(G_{i},P_{i})_{i\in I}$ is said to be a \textit{qualitative
game.}
\end{definition}

\begin{definition}
A \textit{maximal element} for $G$ is defined as a point $x^{\ast }\in
\tprod_{i\in I}G_{i}$ such that for each $i\in I$, $P_{i}(x^{\ast
})=\emptyset $.\medskip
\end{definition}

A qualitative game $G=(G_{i},P_{i})_{i\in I}$ generalizes a strategic game $%
\Gamma =(G_{i},u_{i})_{i\in I}$ by defining

$P_{i}(x)=\{y_{i}\in G_{i}:u_{i}(y_{i},x_{-i})>u_{i}(x,x_{-i})\}.$

\begin{example}
Let $\Gamma =(G_{i},u_{i})_{i\in I}$ be a strategic game, $I=\{1,2\},$ $%
G_{1}=G_{2}=[0,1],$ $u_{i}(x_{1},x_{2})=x_{i}$ $i=1,2.$ The point $x^{\ast
}=(1,1)$ is Nash equilibrium.
\end{example}

The qualitative game corresponding to $\Gamma $ is $G=(G_{i},P_{i})_{i\in
I}, $ where

$P_{1}(x_{1},x_{2})=\{y_{1}\in \lbrack
0,1]:u_{1}(y_{1},x_{2})>u_{1}(x_{1},x_{2})\}=(x_{1},1]$ if $x_{1}\in \lbrack
0,1)$ and $x_{2}\in \lbrack 0,1]$ and

$P_{1}(1,x_{2})=\emptyset $ if $x_{2}\in \lbrack 0,1];$

$P_{2}(x_{1},x_{2})=\{y_{2}\in \lbrack
0,1]:u_{2}(x_{1},y_{2})>u_{2}(x_{1},x_{2})\}=(x_{2},1]$ if $x_{1}\in \lbrack
0,1]$ and $x_{2}\in \lbrack 0,1)$ and

$P_{2}(x_{1},1)=\emptyset $ if $x_{1}\in \lbrack 0,1];$

$(x_{1},x_{2})=(1,1)$ is a maximal element: $P_{1}(1,1)=P_{2}(1,1)=\emptyset
.$\medskip

Now we define a transitivity type of correspondences.

\begin{definition}
Let $\tprod_{i\in I}G_{i}$ be a product space and let $P:\tprod_{i\in
I}G_{i}\rightarrow 2^{G_{i}}$ be a correspondence. We say that $P$ \textit{%
has the property }$T$ if $y_{i}\in P(x)$ implies cl$P(y_{i},x_{-i})\subset
P(x).$
\end{definition}

\begin{example}
$P_{1}$ and $P_{2}$ from Example 1 have the property $T:$
\end{example}

if $y_{1}\in P_{1}(x),$ then $y_{1}\in (x_{1},1]$ and cl$%
P_{1}(y_{1},x_{2})=[y_{1},1]\subset (x_{1},1]$

\begin{definition}
Let $\tprod_{i\in I}G_{i}$ be a product space and let $P,Q:\tprod_{i\in
I}G_{i}\rightarrow 2^{G_{i}}$ be correspondences. We say that the pair $%
(P,Q) $ \textit{has} \textit{the property }$T$ if for each $x\in
\tprod_{i\in I}G_{i},$ $P(x)\subset Q(x)$ and $y_{i}\in P(x)$ implies $%
Q(y_{i},x_{-i})\subset P(x).$
\end{definition}

\begin{example}
If we take $Q=$cl$P$ in Example 2, we obtain that the pair $(P,Q)$ has the
property $T.\medskip $
\end{example}

\begin{example}
Let $P,Q:[0,2]\times \lbrack 0,2]\rightarrow 2^{[0,2]}$ be defined by
\end{example}

$P(x_{1},x_{2})=\left\{ 
\begin{array}{c}
(1,x_{2}]\text{ if }x_{1}\in \lbrack 0,1]\text{ and }x_{2}\in (1,2]; \\ 
\emptyset \text{ otherwise \ \ \ \ \ \ \ \ \ \ \ \ \ \ \ \ \ \ \ \ \ \ \ \ \
\ \ \ \ }%
\end{array}%
\right. $ and

$Q(x_{1},x_{2})=\left\{ 
\begin{array}{c}
\lbrack x_{1},x_{2}]\text{ if }0\leq x_{1}<x_{2}\leq 2; \\ 
\lbrack x_{2},x_{1}]\text{ if }0\leq x_{2}<x_{1}\leq 2; \\ 
\{x_{1}\}\text{ \ if\ \ }0\leq x_{1}=x_{2}\leq 2.%
\end{array}%
\right. $

We have that $P(x_{1},x_{2})\subset Q(x_{1},x_{2})$ for each $%
(x_{1},x_{2})\in \lbrack 0,2]\times \lbrack 0,2]$ and if $y\in
P(x_{1},x_{2}),$ it follows that $x_{1}\in \lbrack 0,1]$, $x_{2}\in (1,2]$
and $y\in (1,x_{2}]$ which implies $Q(y,x_{2})=[y,x_{2}]\subset
(1,x_{2}]=P(x_{1},x_{2}).$

We note that $x_{1}\in Q(x_{1},x_{2})$ for each $(x_{1},x_{2})\in \lbrack
0,2]\times \lbrack 0,2]$ and $Q$ has convex closed values, so that $Q$
verifies the assumptions stated in the hypothesis of the theorems from
Section 3.4

We introduce the following class of games.

Let $I$ be a non-empty set (the set of agents). For each $i\in I$, let $%
G_{i} $ be a non-empty topological vector space representing the set of
actions, and let $P_{i},Q_{i}:\tprod_{i\in I}G_{i}\rightarrow 2^{G_{i}}$ be
correspondences $(P_{i}$ is the preference correspondence).

\begin{definition}
The family $G=(G_{i},P_{i},Q_{i})_{i\in I}$ is said to be a \textit{general} 
\textit{qualitative game with property }$T$ if for each $i\in I,$ the pair $%
(P_{i},Q_{i})$ satisfies the property $T.$\textit{.}
\end{definition}

\begin{definition}
A \textit{maximal element} for the general qualitative game $G$ is defined
as a point $x^{\ast }\in \tprod_{i\in I}G_{i}$ such that for each $i\in I$, $%
P_{i}(x^{\ast })=\emptyset $.\medskip
\end{definition}

\section{GAME\ REDUCTION}

\subsection{DEFINITIONS}

This subsection gives preliminary definitions on parings and strict
dominance for qualitative games.\medskip

We introduce a relation of strict dominance for the qualitative games, which
generalizes the notion of dominance due to Dufwenberg and Stegeman [8].

\begin{definition}
A paring of $G=(G_{i},P_{i})_{i\in I}$ is $H=(H_{i},P_{i|\tprod_{i\in
I}H_{i}})_{i\in I},$ where $H_{i}\subseteq G_{i}.$
\end{definition}

\begin{definition}
Given a pairing $H$ of $G$, the strict dominance relation $\succ _{H}$ on $%
G_{i}$ can be defined$:$
\end{definition}

for $x_{i},y_{i}\in G_{i},$ $y_{i}\succ _{H}x_{i}$ if $H_{-i}\neq \emptyset $
and $y_{i}\in P_{i}(x_{i},x_{-i})$ for each $x_{-i}\in H_{-i},$ which is
equivalent with $H_{-i}\neq \emptyset $ and $y_{i}\in \cap _{x_{-i}\in
H_{-i}}P_{i}(x_{i},x_{-i}).\medskip $

Let us consider parings $G,H$ with the property that $H_{i}\subseteq G_{i}$
for each $i\in I.$ We generalize, for qualitative games, the types of game
reduction used by Dufwenberg and Stegeman [8], in the following way.

\begin{definition}
i) We define the reduction $G\rightarrow H$ if for each $x_{i}\in
G_{i}\backslash H_{i}$, $\cap _{x_{-i}\in G_{-i}}P_{i}(x_{i},x_{-i})\neq
\emptyset .$
\end{definition}

ii) the reduction $G\rightarrow H$ is called fast if $\cap _{x_{-i}\in
G_{-i}}P_{i}(x_{i},x_{-i})\neq \emptyset $ for some $x_{i}\in G_{i}$ implies 
$x_{i}\notin H_{i}.$

iii) the reduction $G\rightarrow ^{\ast }H$ is defined by the existence of
(finite or countable infinite) sequence of parings $R^{t}$ of $G,$ $%
t=0,1,2...$, such that $R^{0}=G,$ $R^{t}\rightarrow R^{t+1}$ fast for each $%
t\geq 0$ and $H_{i}=\cap _{t}R_{i}^{t}$ for each $i\in I;$

iv) $H$ is said to be a maximal $(\rightarrow ^{\ast })$-reduction of $G$ if 
$G\rightarrow ^{\ast }H$ and $H\rightarrow H^{\prime }$ only for $%
H=H^{\prime }.\medskip $

Now, we consider the iterated elimination of strictly dominated strategies
which involves elimination of strategies that are strictly dominated by a
strategy from the currently considered game $H$ and not the initial game $G.$
We study the reductions of qualitative games in the form below and we
introduce the next definition which generalizes the one of Gilboa, Kalai and
Zemel [9] for qualitative games.

\begin{definition}
We say that $G^{\prime }=(H_{1},...,H_{n},P_{1}^{\prime },...P_{n}^{\prime
}) $ is a restriction of a game $G=(G_{1},...G_{n},P_{1},...P_{n})$ if each $%
H_{i}$ is a (possibly empty) subset of $X_{i}$ and $P_{i}^{\prime
}=(P_{i}\cap H_{i})_{|\tprod_{i=1}^{n}H_{i}}.$ If all $H_{i}$ are empty, we
call $H$ an empty restriction.
\end{definition}

\begin{definition}
i) We define the reduction $G\Rightarrow H$ if for each $x_{i}\in
G_{i}\backslash H_{i}$, $\cap _{x_{-i}\in H_{-i}}P_{i}(x_{i},x_{-i})\cap
H_{i}\neq \emptyset .$
\end{definition}

ii) the reduction $G\Rightarrow H$ is called fast if $\cap _{x_{-i}\in
H_{-i}}P_{i}(x_{i},x_{-i})\cap H_{i}\neq \emptyset $ implies $x_{i}\notin
H_{i}.$

iii) the reduction $G\Rightarrow ^{\ast }H$ is defined by the existence of
(finite or countably infinite) sequence of parings $R^{t}$ of $G,$ $%
t=0,1,2...$, such that $R^{0}=G,$ $R^{t}\rightarrow R^{t+1}$ fast for each $%
t\geq 0$ and $H_{i}=\cap _{t}R_{i}^{t}$ for each $i\in I;$

iv) $H$ is said to be a maximal $(\Rightarrow ^{\ast })$-reduction of $G$ if 
$G\Rightarrow ^{\ast }H$ and $H\Rightarrow H^{\prime }$ only for $%
H=H^{\prime }.\medskip $

We further generalize the definition considered by Milgrom and Roberts [13].

\begin{definition}
i) We define the reduction $G\rightarrowtail H$ if for each $x_{i}\in
G_{i}\backslash H_{i}$, $\cap _{x_{-i}\in H_{-i}}P_{i}(x_{i},x_{-i})\neq
\emptyset .$
\end{definition}

ii) the reduction $G\rightarrowtail H$ is called fast if $\cap _{x_{-i}\in
H_{-i}}P_{i}(x_{i},x_{-i})\neq \emptyset $ implies $x_{i}\notin H_{i}.$

iii) the reduction $G\rightarrowtail ^{\ast }H$ is defined by the existence
of (finite or countable infinite) sequence of parings $R^{t}$ of $G,$ $%
t=0,1,2...$, such that $R^{0}=G,$ $R^{t}\rightarrowtail R^{t+1}$ fast for
each $t\geq 0$ and $H_{i}=\cap _{t}R_{i}^{t}$ for each $i\in I;$

iv) $H$ is said to be a maximal $(\rightarrowtail ^{\ast })$-reduction of $G$
if $G\rightarrowtail ^{\ast }H$ and $H\rightarrowtail H^{\prime }$ only for $%
H=H^{\prime }.\medskip $

We have the following lemma.\medskip

\begin{lemma}
If we have $G\rightarrowtail ^{\ast }H$ and $G\Rightarrow ^{\ast }H^{\prime
} $ maximal reductions such that $H_{i},H_{i}^{\prime }\neq \emptyset $ for
each $i\in I,$ then, $H_{i}\subseteq H_{i}^{\prime }$ for each $i\in I.$
\end{lemma}

\textit{Proof. }Let us\textit{\ }assume that\textit{\ }the reduction $%
G\rightarrowtail ^{\ast }H$ is defined by the existence of (finite or
countable infinite) sequence of parings $R^{t}$ of $G,$ $t=0,1,2...$, such
that $R^{0}=G,$ $R^{t}\rightarrowtail R^{t+1}$ fast for each $t\geq 0$ and $%
H_{i}=\cap _{t}R_{i}^{t}$ for each $i\in I.$ Assume also that the reduction $%
G\Rightarrow ^{\ast }H^{\prime }$ is defined by the existence of (finite or
countable infinite) sequence of parings $R^{\prime t}$ of $G,$ $t=0,1,2...$,
such that $R^{\prime 0}=G,$ $R^{\prime t}\rightarrow R^{\prime t+1}$ fast
for each $t\geq 0$ and $H_{i}^{\prime }=\cap _{t}R_{i}^{\prime t}$ for each $%
i\in I.$

Since for each $t,$ the reductions $R^{t}\rightarrowtail R^{t+1}$ and $%
R^{\prime t}\rightarrow R^{\prime t+1}$ are fast, then, for each $t,$ $%
R^{t}\subseteq R^{\prime t}$.

We prove the last assertion by induction. Suppose the claim holds for all $%
t\leq T$ and consider the induction step for $T+1.$ Let $i\in I$ be fixed
and let $x_{i}\in R_{i}^{T+1}$. Then, $x_{i}\in R_{i}^{T}\subseteq
R_{i}^{\prime T}$ and $\cap _{x_{-i}\in
R_{-i}^{T}}P_{i}(x_{i},x_{-i})=\emptyset .$ Since $R^{T}\subseteq R^{\prime
T},$ we have that $\cap _{x_{-i}\in R_{-i}^{\prime
T}}P_{i}(x_{i},x_{-i})=\emptyset $ \ and $\cap _{x_{-i}\in R_{-i}^{\prime
T}}P_{i}(x_{i},x_{-i})\cap R_{i}^{\prime T+1}=\emptyset .$ Consequently, $%
\cap _{x_{-i}\in R_{-i}^{\prime T+1}}P_{i}(x_{i},x_{-i})\cap R_{i}^{\prime
T+1}=\emptyset .$ Hence, $x_{i}\in R_{i}^{\prime T+1}$ and $%
R_{i}^{T+1}\subseteq R_{i}^{\prime T+1}.$

Therefore, $\cap _{t}R_{i}^{t}\subseteq \cap _{t}R_{i}^{\prime t}$ for each $%
i\in I$, that is $H_{i}\subseteq H_{i}^{\prime }$ for each $i\in I.$ $%
\square $

\subsection{THE PROBLEM OF GAME REDUCTION AND ORDER INDEPENDENCE}

We reconsider the iterative processes in which the dominated strategies are
removed and the limit of this process can be an empty or a nonempty set of
strategies. If, for the case of the strategic games, the remained strategies
are the rationalizable ones, it seems that the process itself deserves to be
studied and leads us to a kind of 
\'{}%
rationalizable%
\'{}
result, also for the case of generalized games. In order to better
understand the problem and motivate our demarche, we give several examples.

We first consider example 1 due to Dufwenberg and Stegeman [8] and its
generalization.

\begin{example}
Let $I=\{1,2\},$ $X_{1}=X_{2}=[0,1],$ $u_{i}:G_{i}\times G_{j}\rightarrow 
\mathbb{R},$ where for each $i\in \{1,2\},$ $u_{i}$ is defined in the
following way
\end{example}

$u_{i}(x,y)=x$ if $x\in \lbrack 0,1)$ and $y\in \lbrack 0,1];$

$u_{i}(1,y)=0$ if $y\in \lbrack 0,1);$

$u_{i}(1,1)=1$.

Then,

$P_{1}(x,y)=\{x_{1}\in \lbrack 0,1]:u_{1}(x_{1},y)>u_{1}(x,y)\}=(x,1]$ if $%
x\in \lbrack 0,1),$ $y\in \lbrack 0,1];$

$P_{1}(1,y)=\{x_{1}\in \lbrack 0,1]:u_{1}(x_{1},y)>u_{1}(1,y)\}=[0,1)$ if $%
y\in \lbrack 0,1);$

$P_{1}(1,1)=\{x_{1}\in \lbrack 0,1]:u_{1}(x,1)>u_{1}(1,1)\}=\emptyset .$

$P_{2}(x,y)=\{y_{1}\in \lbrack 0,1]:u_{2}(x,y_{2})>u_{2}(x,y)\}=(y,1]$ if $%
y\in \lbrack 0,1),$ $x\in \lbrack 0,1];$

$P_{2}(x,1)=\{y_{2}\in \lbrack 0,1]:u_{2}(x,y_{2})>u_{2}(x,1)\}=[0,1)$ if $%
x\in \lbrack 0,1);$

$P_{2}(1,1)=\{y_{2}\in \lbrack 0,1]:u_{2}(1,y)>u_{2}(1,1)\}=\emptyset .$

By eliminating $X_{i}\backslash \{1,x\}$ for $i\in \{1,2\}$ and $x\in
\lbrack 0,1),$ we have that

$H_{1}=H_{2}=\{1,x\}$

$P_{1}(1,1)=\emptyset ;$ $P_{1}(1,x)=[0,1);$ $P_{1}(x,1)=(x,1]$ and $%
P_{1}(x,x)=(x,1];$

$P_{2}(1,1)=\emptyset ;$ $P_{2}(x,1)=[0,1);$ $P_{2}(1,x)=(x,1]$ and $%
P_{2}(x,x)=(x,1].$

This game can not be $\rightarrow $ reduced. In this case, since $x$ is
arbitrary, the IESDS procedure is an order dependent one.

\begin{example}
We consider the game from example 2.
\end{example}

If $H_{i}=\{x_{i}\}$ $i\in \{1,2\},$ then $P_{1}^{\prime }(x)=P_{1}(x)\cap
\{x_{1}\}=(x_{1},1]\cap \{x_{1}\}=\emptyset ;$

\ \ \ \ \ \ \ \ \ \ \ \ \ \ \ \ \ \ \ \ \ \ \ \ \ \ \ $P_{2}^{\prime
}(x)=P_{2}(x)\cap \{x_{2}\}=(x_{2},1]\cap \{x_{2}\}=\emptyset .$

It follows that $(x_{1},x_{2})$ is a maximal element for the $\Rightarrow
^{\ast }$reduced game and that it is not a maximal element for the initial
game.

The example of Dufwenberg and Stegeman [8] involves games with discontinuous
utility functions. These authors let an open problem for researcers to find
classes of games for which order independence holds. As it can be seen in
this subsection, the problem of the existence of nonempty maximal reductions
can be reconsidered for qualitative games, which can also be generalizations
of discontinuous strategic games. The aim of this paper is mainly to
generalize the results obtained by Dufwenberg and Stegeman [8] and Apt [1].

\subsection{CONDITIONS $C(t)$ AND $D(t)$}

Let us consider a game reduction from $G$ to $H$, defined by a finite or
countable infinite sequence of parings $R^{t}$ of $G.$

We generalize the property $C(t)$ proposed by Apt [1]. It refers to the fact
that every strictly dominated strategy on each $R^{t}$ has an undominated
dominator.

\begin{definition}
We call Condition $C(t)$ the following property of the initial game: for all 
$i\in I,$
\end{definition}

$\forall $ $x_{i}\in G_{i},$ $(\exists $ $y_{i}\in G_{i},$ $y_{i}\succ
_{R^{t}}x_{i}\Rightarrow \exists x_{i}^{\ast }\in G_{i}$ $($ $x_{i}^{\ast
}\succ _{R^{t}}x_{i}$ $\wedge $ $\nexists $ $y_{i}\in G_{i},$ $y_{i}\succ
_{R^{t}}x_{i}^{\ast })$

which is equivalent with

$\forall $ $x_{i}\in G_{i},$ $R_{-i}^{t}\neq \emptyset $ and $\exists $ $%
y_{i}\in P_{i}(x_{i},x_{-i})$ $\forall x_{-i}\in R_{-i}^{t}\Rightarrow
\exists x_{i}^{\ast }\in G_{i}$ such that $x_{i}^{\ast }\in
P_{i}(x_{i},x_{-i})$ $\forall x_{-i}\in R_{-i}^{t}$ and $\cap _{x_{-i}\in
R_{-i}^{t}}P_{i}(x_{i}^{\ast },x_{-i})=\emptyset .\medskip $

\begin{remark}
For $H=G,$ we obtain the following:
\end{remark}

for all $i\in I,$

$\forall $ $x_{i}\in G_{i},$ $(\exists $ $y_{i}\in G_{i},$ $y_{i}\succ
_{G_{i}}x_{i}\Rightarrow \exists x_{i}^{\ast }\in G_{i}$ $($ $x_{i}^{\ast
}\succ _{G_{i}}x_{i}$ $\wedge $ $\nexists $ $y_{i}\in G_{i},$ $y_{i}\succ
_{G_{i}}x_{i}^{\ast }).\medskip $

The next property concerns the strict dominance in the reduction $R^{t}.$ It
is also an extension of Condition $D(t)$ considered by Apt [1].

\begin{definition}
Condition $D(t)$ states that each strategy $x_{i}$ strictly dominated in $%
R^{t}$ is strictly dominated in $R^{t}$ by some strategy in $R^{t}.$
Formally, we can write that for all $i\in I:$
\end{definition}

$\forall $ $x_{i}\in G_{i},$ $(\exists $ $y_{i}\in G_{i},$ $y_{i}\succ
_{R^{t}}x_{i}\Rightarrow \exists x_{i}^{\ast }\in R_{i}^{t}$ $x_{i}^{\ast
}\succ _{R^{t}}x_{i})$

which is equivalent with

$\forall $ $x_{i}\in G_{i},$ $R_{-i}^{t}\neq \emptyset $ and $\cap
_{x_{-i}\in R_{-i}^{t}}P_{i}(x_{i},x_{-i})\neq \emptyset $ $\Rightarrow \cap
_{x_{-i}\in R_{-i}^{t}}P_{i}(x_{i},x_{-i})\cap R_{i}^{t}\neq \emptyset $.

The next lemma states the relation between the two properties introduced
above.

\begin{lemma}
For each $t,$ the property $C(t)$ implies the property $D(t)$.
\end{lemma}

\textit{Proof.} Let us consider $i\in I.$ Assume that $x_{i}\in G_{i},$ $%
x_{i}^{\prime }\in G_{i},$ $x_{i}^{\prime }\in P_{i}(x_{i},x_{-i})$ $\forall
x_{-i}\in R_{-i}^{t}.$ The Condition $C(t)$ implies that there exists $%
x_{i}^{\ast }\in G_{i}$ such that $x_{i}^{\ast }\in P_{i}(x_{i},x_{-i})$ $%
\forall x_{-i}\in R_{i}^{t}$ and $\cap _{x_{-i}\in
R_{-i}^{t}}P_{i}(x_{i}^{\ast },x_{-i})=\emptyset .$ For each $s\leq t,$ $%
R_{i}^{t}\subseteq R_{i}^{s},$ then for each $s\leq t,$ $\cap _{x_{-i}\in
R_{-i}^{s}}P_{i}(x_{i}^{\ast },x_{-i})=\emptyset .$ It follows that $%
x_{i}^{\ast }$ is a strategy of player $i$ in all restrictions $R^{s}$ for $%
s\leq t,$ that is $x_{i}^{\ast }\in \cap _{s\leq t}R_{i}^{s}.$ Hence, $%
x_{i}^{\ast }\in R_{i}^{t}$ and we can conclude that $x_{i}^{\ast }\in
P_{i}(x_{i},x_{-i})\cap R_{i}^{t}$ $\forall x_{-i}\in R_{-i}^{t}.$ $\square $

\begin{theorem}
If the property $D(t)$ is fulfilled for each $t$, then a non-empty maximal $%
(\rightarrow ^{\ast })$ reduction of $G$ is the unique maximal $(\rightarrow
^{\ast })$ reduction of $G,$ if the maximal reduction is defined by a finite
sequence of parings.
\end{theorem}

\textit{Proof. Let }$M$ and $M^{\prime }$ be maximal $(\rightarrow ^{\ast
})- $ reductions of $G,$ $M$ being nonempty. Let us consider $G\rightarrow
^{\ast }M^{\prime }$ and $R^{t},$ $t=0,1,2,...,T_{0}$ be the implied finite
sequence of parings. If $M_{i}\nsubseteq M_{i}^{\prime }$ for some $i,$ it
follows that $M_{i}\nsubseteq R_{i}^{t},$ $\forall t>T$ for the largest $T$
such that $R_{i}^{T+1}$ is well-defined and $M_{i}\subseteq R_{i}^{T}$ $%
\forall i\in I.$ Let us take $x_{i}\in M_{i}\backslash R_{i}^{T+1}$ for an $%
i $ fixed. We have that $x_{i}\in R_{i}^{T}\backslash R_{i}^{T+1}$, so that
there exists $y_{i}\in R_{i}^{T}$ such that $y_{i}\in \cap _{x_{-i}\in
R_{i}^{T}}P_{i}(x_{i},x_{-i}).$ Since, in addition, $\emptyset \neq
M_{i}\subseteq R_{i}^{T}$ for each $i\in I,$ it follows that $y_{i}\in \cap
_{x_{-i}\in R_{i}^{T}}P_{i}(x_{i},x_{-i}).$ Let $M=R^{T_{0}}.$ According to
property $D(T_{0}),$ there exists $z_{i}^{\ast }\in M_{i}$ such that $%
z_{i}^{\ast }\in \cap _{x_{-i}\in M_{-i}}P_{i}(x_{i},x_{-i}),$ which
contradicts the fact that $M$ is a maximal $(\rightarrow ^{\ast })-$%
reduction. It remains that $M_{i}\subseteq M_{i}^{\prime }$ for each $i\in I$
and therefore $M^{\prime }$ is nonempty. We can prove in the same way that $%
M_{i}^{\prime }\subseteq M_{i}$ for each $i\in I,$ implying $M=M^{\prime }.$ 
$\square $ $\medskip $

The next theorem states the relation between $(\rightarrowtail ^{\ast })$
and $(\Rightarrow ^{\ast })-$ reductions of a game.

\begin{theorem}
Assume that\textit{\ }the reduction $G\rightarrowtail ^{\ast }H$ is defined
by the existence of (finite or countable infinite) sequence of parings $%
R^{t} $ of $G,$ $t=0,1,2...$, and the reduction $G\Rightarrow ^{\ast
}H^{\prime }$ is defined by the existence of (finite or countable infinite)
sequence of parings $R^{\prime t}$ of $G,$ $t=0,1,2...$
\end{theorem}

\textit{Assume that property }$\mathit{D(}s)$\textit{\ holds for each }$s<t.$%
\textit{\ Then, }$R^{t}=R^{\prime t}.$\textit{\ In particular, if }$\mathit{%
D(}t)$\textit{\ holds for each }$t,$\textit{\ then, }$H=H^{\prime };$ 
\textit{then, }$G$\textit{\ has a unique }$\mathit{(}\Rightarrow ^{\ast })-$%
\textit{maximal reduction, if }$G$\textit{\ has a unique }$(\rightarrowtail
^{\ast })-$\textit{maximal reduction.}

\textit{Proof. }We will first prove\textit{\ }$R_{i}^{\prime t}\subseteq
R_{i}^{t}$ by induction. Suppose that D($s)$ for each $s<t+1.$

Let $x_{i}\in R_{i}^{\prime t+1}.$ Since $R^{\prime t}\rightarrow R^{\prime
t+1}$ is fast, $x_{i}\in R^{\prime t}$ and $\nexists x_{i}^{\prime }\in
R_{i}^{\prime t}$ such that $x_{i}^{\prime }\succ _{R^{\prime t}}x_{i}$.
According to property $D(t),$ $\nexists x_{i}^{\prime }\in G_{i}$ such that $%
x_{i}^{\prime }\succ _{R^{\prime t}}x_{i}$. By the induction hypothesis, $%
R_{i}^{\prime t}\subseteq R_{i}^{t},$ therefore, $\nexists x_{i}^{\prime
}\in G_{i}$ such that $x_{i}^{\prime }\succ _{R^{t}}x_{i}$ and $x_{i}\in
R_{i}^{t}.$ We conclude that $x_{i}$ is a strategy in $R_{i}^{t+1}$ and
then, $R_{i}^{\prime t+1}\subseteq R_{i}^{t+1}.$ We also have from Lemma 1
that $R_{i}^{t+1}\subseteq R_{i}^{\prime t+1}$ and it follows that $%
R_{i}^{t+1}=R_{i}^{\prime t+1}.$ $\square $

\subsection{THE\ MAIN\ RESULTS}

We state the results concerning the iterated elimination of strictly
dominated strategies for several classes of qualitative games. The existence
and uniqueness of maximal reductions are proved.\medskip

Theorem 4 generalizes Lemma of Dufwenberg and Stegeman [8] by considering
qualitative games with $U-$majorized correspondences defined by Yuan and
Tarafdar. A proof of uniqueness of the maximal reduction of $G$ is given in
this case.

We first present the notion of $U$-majorized correspondence, which
generalize the classical upper semicontinuous correspondences.

\begin{definition}
(Yuan and Tarafdar, [20]).\textbf{\ }Let $X$ be a topological space and $Y$
be a non-empty subset of a vector space $E,$ $\theta :X\rightarrow E$ a
function and $\,P:X\rightarrow 2^{Y}$ a correspondence.
\end{definition}

1) $P$ is \textit{of class }$U_{\theta }$ (or $U$) if:

\ \ \ \ \ \ \ i) for each $x\in X$, $\theta (x)\notin P(x)$ and

\ \ \ \ \ \ \ ii) $P$ is upper semicontinuous with closed convex values in $%
Y $;

2) A correspondence $P_{x}:X\rightarrow 2^{Y}$ is a $U_{\theta }$\textit{%
-majorant of }$P$\textit{\ at }$x$ \thinspace if there exists an open
neighborhood $N(x)$ of $x$ such that

\ \ \ \ \ \ \ i)\ for each $z\in V(x)$, $P(z)\subset P_{x}(z)$ and $\theta
(z)\notin P_{x}(z);$

\ \ \ \ \ \ \ \ ii)\ $P_{x}$ is upper semicontinuous with closed convex
values;

3) $P$ is $U_{\theta }-$\textit{majorized }if for each $x\in X$ with $%
P(x)\neq \emptyset ,$ there exists a $U$-majorant $P_{x}$ of $P$ at $x$%
.\smallskip

The following theorem is Ding's result on the existence of maximal elements
of qualitative games. The correspondences are $U-$majorized.

\begin{theorem}
(Ding, [7])Let $X$ be a nonempty subset of a Hausdorff locally convex
topological vector space and $D$ a non-empty compact subset of $X.$ Let $%
P:X\rightarrow 2^{D}$ be a $U-$majorized correspondence. Then, there exists $%
x^{\ast }\in $co$D$ such that $P(x^{\ast })=\emptyset .\medskip $
\end{theorem}

\begin{theorem}
Let $I$ be a set of players, $G_{i}$ be a compact convex subset of a
Hausdorff locally convex topological vector space $\forall i\in I$ and the
general qualitative game $G=(G_{i},P_{i},Q_{i})_{i\in I}$ which satisfies
the property $T$, where $P_{i},Q_{i}:\tprod_{i\in I}G_{i}\rightarrow
2^{G_{i}}$ satisfies the following assumptions for each $i\in I:$
\end{theorem}

\textit{i) }$y_{i}\in Q_{i}(y_{i},x_{-i})$\textit{\ for each }$x_{-i}\in
G_{-i};$

\textit{ii) }$Q_{i}$\textit{\ has convex closed values;}

\textit{iii) }$P_{i}(.,x_{-i})$\textit{\ is }$U-$\textit{majorized on }$%
G_{i} $\textit{\ for each }$x_{-i}\in G_{-i}$\textit{.}

\textit{Then,}

\textit{a) If }$G\rightarrow ^{\ast }H$\textit{\ is a game reduction and if
there exists }$y_{i}\succ _{H}x_{i},$\textit{\ for some }$x_{i},y_{i}\in
G_{i}$\textit{\ and }$i\in I,$\textit{\ there exists }$x_{i}^{\ast }\in
H_{i} $\textit{\ such that }$z_{i}\nsucc _{H}x_{i}^{\ast }\succ _{H}x_{i}$%
\textit{\ }$\forall z_{i}\in G_{i};$

\textit{b) a non-empty maximal }$(\rightarrow ^{\ast })$\textit{\ reduction
of }$G$\textit{\ is the unique maximal }$(\rightarrow ^{\ast })$\textit{\
reduction of }$G.$

\textit{Proof. }a)\textit{\ Let }$R^{t}$ be the sequence of parings of $G,$ $%
t=0,1,2...$, such that $R^{0}=G,$ $R^{t}\rightarrow R^{t+1}$ for each $t\geq
0$ and $H_{i}=\cap _{t}R_{i}^{t}$ for each $i\in I.$ Assume that there
exists $y_{i}\succ _{H}x_{i}.$ Let $Z_{i}=\cap _{x_{-i}\in
H_{-i}}Q(y_{i},x_{-i}).$ According to i), we have that $Z_{i}\neq \emptyset
. $ The set $Z_{i}$ is convex and closed, so it is compact. Since $%
y_{i}\succ _{H}x_{i},$ we have that $H_{-i}\neq \emptyset .$ Let $%
x_{-i}^{\ast }\in H_{-i}$ be fixed and $F_{i}:Z_{i}\times \{x_{-i}^{\ast
}\}\rightarrow 2^{Z_{i}},$ $F_{i}=P_{i|Z_{i}\times \{x_{-i}^{\ast }\}}.$
According to Ding's Theorem, there exists $x_{i}^{\ast }\in Z_{i}$ such that 
$P_{i}(x_{i}^{\ast },x_{-i}^{\ast })=\emptyset .$

We have that $x_{i}^{\ast }\in Q_{i}(y_{i},x_{-i})$ for each $x_{-i}\in
H_{-i}$. The relation $y_{i}\succ _{H}x_{i}$ implies that $y_{i}\in
P_{i}(x_{i},x_{-i})$ $\forall x_{-i}\in H_{-i}$ and since the pair $%
(P_{i},Q_{i})$ has the property $T$ on $\tprod_{i\in I}H_{i}$, it follows
that $x_{i}^{\ast }\in P_{i}(x_{i},x_{-i})$ $\forall x_{-i}\in H_{-i}.$

If there exists $z_{i}\in P_{i}(x_{i}^{\ast },x_{-i})\subset
Q_{i}(x_{i}^{\ast },x_{-i})$ $\forall x_{-i}\in H_{-i},$ then $z_{i}\in
Z_{i} $ and $z_{i}\in P_{i}(x_{i}^{\ast },x_{-i}^{\ast }),$ which is a
contradiction.

Since $H_{-i}\subseteq R_{-i}^{t},$ $\cap _{x_{-i}\in H_{-i}}P(x_{i}^{\ast
},x_{-i})=\emptyset $ and then, $\cap _{x_{-i}\in R_{-i}^{t}}P(x_{i}^{\ast
},x_{-i})=\emptyset .$ We conclude that $x_{i}^{\ast }\in R_{i}^{t}$ $%
\forall t\geq 0$ and $x_{i}^{\ast }\in H_{i}.$

b) \textit{Let }$M$ and $M^{\prime }$ be maximal $(\rightarrow ^{\ast })-$
reductions of $G,$ $M$ being nonempty. Let us consider $G\rightarrow ^{\ast
}M^{\prime }$ and $R^{t},$ t=0,1,2..., be the implied finite or infinite
sequence of parings. If $M_{i}\nsubseteq M_{i}^{\prime }$ for some $i,$ it
follows that $M_{i}\nsubseteq R_{i}^{t},$ $\forall t>T$ for the largest $T$
such that $R_{i}^{T+1}$ is well-defined and $M_{i}\subseteq R_{i}^{T}$ $%
\forall i\in I.$ Let us take $x_{i}\in M_{i}\backslash R_{i}^{T+1}$ for an $%
i $ fixed. We have that $x_{i}\in R_{i}^{T}\backslash R_{i}^{T+1}$, so that
there exists $y_{i}\in R_{i}^{T}$ such that $y_{i}\in \cap _{x_{-i}\in
R_{-i}^{T}}P_{i}(x_{i},x_{-i}).$ Since, in addition, $\emptyset \neq
M_{i}\subseteq R_{i}^{T}$ for each $i\in I,$ it follows that $y_{i}\in \cap
_{x_{-i}\in M_{-i}}P_{i}(x_{i},x_{-i}).$ According to i) there exists $%
z_{i}^{\ast }\in M_{i}$ such that $z_{i}^{\ast }\in \cap _{x_{-i}\in
M_{-i}}P_{i}(x_{i},x_{-i}),$ which contradicts the fact that $M$ is a
maximal $(\rightarrow ^{\ast })-$reduction. It remains that $M_{i}\subseteq
M_{i}^{\prime }$ for each $i\in I$ and therefore $M^{\prime }$ is nonempty.
We can prove that $M_{i}^{\prime }\subseteq M_{i}$ for each $i\in I,$
implying $M=M^{\prime }.$ $\square \medskip $

We obtain the following corollary for qualitative games having upper
semicontinuous correspondences $P_{i},$ $i\in I.$

\begin{corollary}
Let $I$ be a set of players, $G_{i}$ be a compact convex subset of a
Hausdorff locally convex topological vector space $\forall i\in I$ and the
general qualitative game $G=(G_{i},P_{i},Q_{i})_{i\in I}$ which satisfies
the property $T$, where $P_{i},Q_{i}:\tprod_{i\in I}G_{i}\rightarrow
2^{G_{i}}$ satisfies the following assumptions for each $i\in I:$
\end{corollary}

\textit{i) }$y_{i}\in Q_{i}(y_{i},x_{-i})$\textit{\ for each }$x_{-i}\in
G_{-i};$

\textit{ii) }$Q_{i}$\textit{\ has convex closed values;}

\textit{iii) for each }$x\in G_{i}$\textit{, }$x_{i}\notin P_{i}(x)$\textit{%
\ and}

\textit{iv) }$P_{i}$\textit{\ is upper semicontinuous with closed convex
values in }$G_{i}$\textit{;}

\textit{Then,}

\textit{a) If }$G\rightarrow ^{\ast }H$\textit{\ is a game reduction and if
there exists }$y_{i}\succ _{H}x_{i},$\textit{\ for some }$x_{i},y_{i}\in
G_{i}$\textit{\ and }$i\in I,$\textit{\ there exists }$x_{i}^{\ast }\in
H_{i} $\textit{\ such that }$z_{i}\nsucc _{H}x_{i}^{\ast }\succ _{H}x_{i}$%
\textit{\ }$\forall z_{i}\in G_{i};$

\textit{b) a non-empty maximal }$(\rightarrow ^{\ast })$\textit{\ reduction
of }$G$\textit{\ is the unique maximal }$(\rightarrow ^{\ast })$\textit{\
reduction of }$G.$\medskip

Liu and Cai [11] defined the correspondences of class $Q$ and the $Q$%
-majorized correspondences.\medskip

\begin{definition}
(Liu and Cai, [11]) Let $X$ be a topological space and $Y$ be a non-empty
subset of a vector space $E$, $\theta :X\rightarrow E$ a function and $%
\,P:X\rightarrow 2^{Y}$ a correspondence.{}
\end{definition}

1) $P$ \textit{is of class }$Q_{\theta }$ (or $Q$) if:

\qquad i) for each $x\in X$, $\theta (x)\notin $cl$P(x)$ and

\qquad ii) $P$ is lower semicontinuous with open and convex values in $Y$;

2) A correspondence $P_{x}:X\rightarrow 2^{Y}$ is a $Q_{\theta }$\textit{%
-majorant of }$P$ at $x,$ \thinspace if there exists an open neighborhood $%
N(x)$ of $x$ such that:

\qquad i) for each $z\in N(x)$, $P(z)\subset P_{x}(z)$ and $\theta (z)\notin 
$cl$P_{x}(z);$

\qquad ii) $P_{x}$ is lower semicontinuous with open convex values;

3) $P$\textit{\ is }$Q_{\theta }$\textit{-majorized} if for each $x\in X$
with $P(x)\neq \emptyset ,$ there exists a $Q_{\theta }$-majorant $P_{x}$ of 
$P$ at $x$.\medskip

The next result is also due to Liu and Cai and states the maximal element
existence for qualitative games with $Q_{\theta }-$majorized correspondences.

\begin{theorem}
(Liu, Cai, [11]). Let $X$ be a convex paracompact subset of a locally convex
Hausdorff topological vector space $E,$ let $D$ be a nonempty compact
metrizable subset of $X.$ Let $P:X\rightarrow 2^{D}$ be a $Q_{\theta }-$%
majorized correspondence. Then, there exists $x^{\ast }\in X$ such that $%
P(x^{\ast })=\emptyset .\medskip $
\end{theorem}

Theorem 6 concerns the games with $Q_{\theta }-$majorized correspondences.

\begin{theorem}
Let $I$ be a set of players, $G_{i}$ be a compact convex subset of a
Hausdorff locally convex topological vector space $\forall i\in I$ and the
general qualitative game $G=(G_{i},P_{i},Q_{i})_{i\in I}$ which satisfies
the property $T$, where $P_{i},Q_{i}:\tprod_{i\in I}G_{i}\rightarrow
2^{G_{i}}$ satisfy the following assumptions for each $i\in I:$
\end{theorem}

\textit{i) }$y_{i}\in Q_{i}(y_{i},x_{-i})$\textit{\ for each }$x_{-i}\in
G_{-i};$

\textit{ii) }$Q_{i}$\textit{\ has convex closed values;}

\textit{iii) }$P_{i}(.,x_{-i})$\textit{\ is }$Q_{\theta }-$\textit{majorized
on }$G_{i}$\textit{\ for each }$x_{-i}\in G_{-i}$\textit{.}

\textit{Then,}

\textit{a) If }$G\rightarrow ^{\ast }H$\textit{\ is a a game reduction and
if there exists }$y_{i}\succ _{H}x_{i},$\textit{\ for some }$x_{i},y_{i}\in
G_{i}$\textit{\ and }$i\in I,$\textit{\ there exists }$x_{i}^{\ast }\in
H_{i} $\textit{\ such that }$z_{i}\nsucc _{H}x_{i}^{\ast }\succ _{H}x_{i}$%
\textit{\ }$\forall z_{i}\in G_{i};$

\textit{b) a non-empty maximal }$\rightarrow ^{\ast }$\textit{reduction of }$%
G$\textit{\ is the unique maximal reduction of }$G.\medskip $

Since a\textit{\ }correspondence of class\textit{\ }$Q$ is $Q$-majorized$,$
we obtain the following corollary.

\begin{corollary}
Let $I$ be a set of players, $G_{i}$ be a compact convex subset of a
Hausdorff locally convex topological vector space $\forall i\in I$ and the
general qualitative game $G=(G_{i},P_{i},Q_{i})_{i\in I}$ which satisfies
the property $T$, where $P_{i},Q_{i}:\tprod_{i\in I}G_{i}\rightarrow
2^{G_{i}}$ satisfy the following assumptions for each $i\in I:$
\end{corollary}

\textit{i) }$y_{i}\in Q_{i}(y_{i},x_{-i})$\textit{\ for each }$x_{-i}\in
G_{-i};$

\textit{ii) }$Q_{i}$\textit{\ has convex closed values;}

\textit{iii) for each }$x_{i}\in G_{i}$\textit{, }$x_{i}\notin $\textit{cl}$%
P_{i}(x)$\textit{\ and}

\textit{iv) }$P_{i}$\textit{\ is lower semicontinuous with open and convex
values in }$G_{i}$\textit{;}

\textit{Then,}

\textit{a) If }$G\rightarrow ^{\ast }H$\textit{\ is a a game reduction and
if there exists }$y_{i}\succ _{H}x_{i},$\textit{\ for some }$x_{i},y_{i}\in
G_{i}$\textit{\ and }$i\in I,$\textit{\ there exists }$x_{i}^{\ast }\in
H_{i} $\textit{\ such that }$z_{i}\nsucc _{H}x_{i}^{\ast }\succ _{H}x_{i}$%
\textit{\ }$\forall z_{i}\in G_{i};$

\textit{b) a non-empty maximal }$\rightarrow ^{\ast }$\textit{reduction of }$%
G$\textit{\ is the unique maximal reduction of }$G.\medskip $

We present here the notion of $L_{S}-$majorized correspondence.

Let $X$ be a topological space and $I$ be an index set. For each $i\in I,$
let $Y_{i}$ be a non-empty convex subset of a topological vector space $%
E_{i}.$ Let $Y=\tprod\limits_{i\in I}Y_{i}$ and $S:Y\rightarrow X$ be a
function. For each $i\in I,$ let $P_{i}:X\rightarrow 2^{Y_{i}}$ be a
correspondence.

\begin{definition}
\ (Yuan, [19]).\textit{\ }$P_{i}$ is said to be:
\end{definition}

1) \textit{of class} $L_{S}$ if:

\qquad i) $P_{i}$ has convex values;

\qquad ii) $y_{i}\notin P_{i}(S(y))$ for each $y\in Y$, and

\qquad iii) $P_{i}^{-1}(y_{i})=\{x\in X:y_{i}\in P_{i}(x)\}$ is open in $X$
for each $y_{i}\in Y_{i}.$

2) $L_{S}$-\textit{majorized }if for each $x\in X,$ there exists an open
neighborhood $N(x)$ of $x$ in $X$ and a correspondence with convex values $%
B_{x}:X\rightarrow 2^{Y_{i}}$ such that:

\qquad i) \ $P_{i}(z)\subset B_{x}(z)$ for each $z\in N(x);$

\qquad ii) $y_{i}\notin B_{x}(S(y))$ for each $y\in Y$ and

\qquad iii) $B_{x}^{-1}(y_{i})$ is open in $X$ for each $y_{i}\in
Y_{i}.\medskip $

We also have the following theorem due to Chang [4] on non-compact spaces.

\begin{theorem}
(Chang, [4]) Let $X$ be a convex subset of a Hausdorff topological vector
space $E$ and let $P:X\rightarrow 2^{X}$ be a $L_{S}-$majorized
correspondence. Suppose that there exists a compact set $D$ in $X$ such
that, for each finite subset $S$ of $X,$ there exists a convex compact set $%
K,$ which contains $S$ and which satisfies $K\backslash \cup _{x\in
K}P^{-1}(x)\subset D.$ Then, there exists $x^{\ast }\in D$ such that $%
P(x^{\ast })=\emptyset .\medskip $
\end{theorem}

The following result is a consequence of Chang's Theorem.

\begin{theorem}
Let $I$ be a set of players, $G_{i}$ be a compact convex subset of a
Hausdorff locally convex topological vector space $\forall i\in I$ and the
general qualitative game $G=(G_{i},P_{i},Q_{i})_{i\in I},$ which satisfies
the property $T,$ where $P_{i},Q_{i}:\tprod_{i\in I}G_{i}\rightarrow
2^{G_{i}}$ satisfy the following assumptions for each $i\in I:$
\end{theorem}

\textit{i) }$y_{i}\in Q_{i}(y_{i},x_{-i})$\textit{\ for each }$x_{-i}\in
G_{-i};$

\textit{ii) }$Q_{i}$\textit{\ has convex closed values;}

\textit{iii) }$P_{i}(.,x_{-i})$\textit{\ is }$L_{S}-$\textit{majorized on }$%
G_{i}$\textit{\ for each }$x_{-i}\in G_{-i}$\textit{.}

\textit{Then,}

\textit{a) If }$G\rightarrow ^{\ast }H$\textit{\ is a a game reduction and
if there exists }$y_{i}\succ _{H}x_{i},$\textit{\ for some }$x_{i},y_{i}\in
G_{i}$\textit{\ and }$i\in I,$\textit{\ there exists }$x_{i}^{\ast }\in
H_{i} $\textit{\ such that }$z_{i}\nsucc _{H}x_{i}^{\ast }\succ _{H}x_{i}$%
\textit{\ }$\forall z_{i}\in G_{i};$

\textit{b) a non-empty maximal }$\rightarrow ^{\ast }$\textit{reduction of }$%
G$\textit{\ is the unique maximal reduction of }$G.\medskip $

Since a\textit{\ }correspondence of class\textit{\ }$L$ is $L_{S}$-majorized$%
,$ we obtain the following corollary.

\begin{corollary}
Let $I$ be a set of players, $G_{i}$ be a compact convex subset of a
Hausdorff locally convex topological vector space $\forall i\in I$ and the
general qualitative game $G=(G_{i},P_{i},Q_{i})_{i\in I},$ which satisfies
the property $T,$ where $P_{i},Q_{i}:\tprod_{i\in I}G_{i}\rightarrow
2^{G_{i}}$ satisfy the following assumptions for each $i\in I:$
\end{corollary}

\textit{i) }$y_{i}\in Q_{i}(y_{i},x_{-i})$\textit{\ for each }$x_{-i}\in
G_{-i};$

\textit{ii) }$Q_{i}$\textit{\ has convex closed values;}

\textit{iii) }$P_{i}$\textit{\ has convex values;}

\textit{iv) }$x_{i}\notin P_{i}(x)$\textit{\ for each }$x\in \tprod_{i\in
I}G_{i}$\textit{;}

\textit{v) }$P_{i}^{-1}(y_{i})=\{x\in \tprod_{i\in I}G_{i}:y_{i}\in
P_{i}(x)\}$\textit{\ is open in }$\tprod_{i\in I}G_{i}$\textit{\ for each }$%
y_{i}\in G_{i}.$

\textit{Then,}

\textit{a) If }$G\rightarrow ^{\ast }H$\textit{\ is a a game reduction and
if there exists }$y_{i}\succ _{H}x_{i},$\textit{\ for some }$x_{i},y_{i}\in
G_{i}$\textit{\ and }$i\in I,$\textit{\ there exists }$x_{i}^{\ast }\in
H_{i} $\textit{\ such that }$z_{i}\nsucc _{H}x_{i}^{\ast }\succ _{H}x_{i}$%
\textit{\ }$\forall z_{i}\in G_{i};$

\textit{b) a non-empty maximal }$\rightarrow ^{\ast }$\textit{reduction of }$%
G$\textit{\ is the unique maximal reduction of }$G.\medskip $

\begin{remark}
For $H=G,$ we obtain that under the conditions of Corollary 3, we have the
following:
\end{remark}

if there exists $y_{i}\succ _{G}x_{i},$\ for some $x_{i},y_{i}\in G_{i}$\
and $i\in I,$\ there exists $x_{i}^{\ast }\in G_{i}$\ such that $z_{i}\nsucc
_{G}x_{i}^{\ast }\succ _{G}x_{i}$\ $\forall z_{i}\in G_{i};\medskip $

The next theorem is a generalization of Theorem 1 of Dufwenberg and Stegeman
[8] for the class of qualitative games.

\begin{theorem}
Let $G=(G_{i},P_{i},Q_{i})_{i\in I}$ be a general qualitative game, where
for each $i\in I,$ $G_{i}$ is a non-empty compact subset of a Hausdorff
topological vector space and $P_{i}:\tprod_{i\in I}G_{i}$ such that:
\end{theorem}

\textit{i) }$P_{i}$\textit{\ has compactly open lower sections and convex
values;}

\textit{ii) }$x_{i}\notin P_{i}(x)$\textit{\ }$\forall x\in \tprod_{i\in
I}G_{i};$

\textit{iii) }$y_{i}\in Q_{i}(y_{i},x_{-i})$\textit{\ for each }$x_{-i}\in
G_{-i};$

\textit{iv) }$Q_{i}$\textit{\ has convex closed values;}

\textit{Then, }$G$\textit{\ has a unique maximal }$\rightarrow ^{\ast }$%
\textit{reduction }$M.$\textit{\ Further, }$\forall i\in I,$\textit{\ }$%
M_{i} $\textit{\ is nonempty, compact and }$P_{i|M}$\textit{\ has compactly
open lower sections.}

\textit{Proof. }We will\textit{\ }prove further\textit{\ }that\textit{\ }$G$
has a nonempty maximal reduction.The proof of its uniqueness is a
consequence of Corollary 3.

1) We first establish that if $G\rightarrow H$ is fast and $H$ is compact, $%
H_{i}$ is compact and nonempty. Choose $i$ such that $H_{i}\neq G_{i}.$
Then, $y_{i}\succ _{G}x$ for some $x,y\in G_{i}$ and then the set $H_{i}$ is
nonempty.

We will prove that $H_{i}$ is compact. Let $y_{i}\in H_{i}$ and $%
Z(y_{i})=C_{X}P_{i}^{-1}(y_{i}),$ where $X=\tprod_{i\in I}G_{i}.$ According
to the assumption ii), $Z(y_{i})\neq \emptyset $ and according to i), $%
Z(y_{i})$ is closed in the compact set $\tprod_{i\in I}G_{i},$ and therefore
it is compact.

Define $Z_{i}(y_{i}):=$pr$_{i}Z(y_{i}).$ We have that $Z(y_{i})$ is a
non-empty closed set in $G_{i}$ with $y_{i}\in Z_{i}(y_{i}).$

Now, we prove that $H_{i}=\cap _{y_{i}\in H_{i}}Z_{i}(y_{i}).$ In order to
show that $H_{i}\subseteq \cap _{y_{i}\in H_{i}}Z_{i}(y_{i}),$ we consider $%
z_{i}\in G_{i}.$ If for every $y_{i}\in H_{i},$ $z_{i}\notin Z(y_{i}),$ it
follows that $y_{i}\in P_{i}(z_{i},x_{-i})$ $\forall x_{-i}\in G_{-i},$ then 
$z_{i}\notin H_{i}$ and therefore $C_{G_{i}}(\cap _{y_{i}\in
H_{i}}Z_{i}(y_{i}))\subset C_{G_{i}}H_{i},$ which implies $H_{i}\subseteq
\cap _{y_{i}\in H_{i}}Z_{i}(y_{i}).$ Now, we want to show that $\cap
_{y_{i}\in H_{i}}Z_{i}(y_{i})\subset H_{i},$ that is $C_{G_{i}}H_{i}%
\subseteq C_{G_{i}}(\cap _{y_{i}\in H_{i}}Z_{i}(y_{i})).$

If $z_{i}\notin H_{i},$ then there exists $x_{i}\in G_{i}$ such that $%
x_{i}\in P_{i}(z_{i},x_{-i})$ $\forall x_{-i}\in G_{-i}.$

According to Remark 2, there exists $x_{i}^{\ast }\in G_{i}$ such that $%
x_{i}^{\ast }\in P_{i}(z_{i},x_{-i})$ $\forall x_{-i}\in G_{-i}.$ It follows
that $z_{i}\notin Z_{i}(x_{i}^{\ast }),$ therefore $z_{i}\notin \cap
_{y_{i}\in X_{i}}Z_{i}(y_{i}),$ and, hence, $C_{G_{i}}H_{i}\subseteq
C_{G_{i}}(\cap _{y_{i}\in H_{i}}Z_{i}(y_{i})).$ Since $H_{i}=\cap _{y_{i}\in
Y_{i}}Z_{i}(y_{i}),$ $H_{i}$ is closed in $G_{i}$ and compact.

2) Let $R^{t},$ $t=0,1,...$ denote the unique sequence of games of $G$ such
that $R^{0}=G$ and $R^{t}\rightarrow R^{t+1}$ is fast $\forall t.$ Result 1
implies that $R^{t}$ is compact and nonempty $\forall t,$ so that $%
M_{i}=\cap _{t}R_{i}^{t}$ is compact and nonempty. According to i), it
follows that $P_{i}^{-1}(y_{i})$ is open in $M_{i}.$ We still have to show
that $M$ is a maximal $\rightarrow ^{\ast }$reduction of $G.$

Let's consider $i\in I$ and $x_{i}\in M_{i}.$ We will prove that $x_{i}$ is
not dominated by any $y_{i}\in M_{i}$. Let $y_{i}\in M_{i},$and let $%
A=C_{X_{-i}}B,$ where $B=\{x_{-i}\in G_{-i}:y_{i}\in P_{i}(x_{i},x_{-i})\}.$
Then $A=C_{X_{-i}}\{x_{-i}\in X_{-i}:(x_{i},x_{-i})\in P_{i}^{-1}(y_{i})\}.$

If $A\cap R_{-i}^{t}=\emptyset $ for every $t$ such that $R^{t}\neq M,$ then 
$y_{i}\in P_{i}(x_{i},x_{-i})$ for each $x_{-i}\in R_{-i}^{t},$ that is $%
y_{i}\succ _{R^{t}}x_{i}$ , which contradicts $x_{i}\in M_{i}.$ Therefore $%
A\cap R_{-i}^{t}$ is nonempty and compact for every $t,$ such that $%
R^{t}\neq M.$ This fact implies $A\cap M_{-i}$ is nonempty and then, $%
y_{i}\notin P_{i}(x_{i},x_{-i})$ $\forall x_{-i}\in M_{-i},$ that is $%
y_{i}\nsucc _{M}x_{i}.$ $\square $

\subsection{MAXIMAL ELEMENTS FOR QUALITATIVE GAMES}

This subsection is meant to prove that the set of maximal elements is
preserved in any game by the process of iterated elimination of strictly
dominated strategies.\medskip

\begin{theorem}
Let $G=(G_{i},P_{i},Q_{i})_{i\in I}$ be a general qualitative game which
satisfies property $T$ and $x_{i}\notin P_{i}(x_{i},x_{-i})$ for each $%
x_{-i}\in \tprod_{i\in I}G_{i}.$ Let us assume that for each $x\in
\tprod_{i\in I}G_{i},$ there exists $z^{\ast }\in \tprod_{i\in I}G_{i}$ such
that $z_{i}^{\ast }\in Q_{i}(z_{i},x_{-i})$ for all $z\in \tprod_{i\in
I}G_{i}$ and $i\in I.$ If $H$ is a $(\Rightarrow ^{\ast })-$reduction of $G,$
then games $G$ and $H$ have the same maximal elements.
\end{theorem}

\textit{Proof. }Let $R^{t},$ $t=0,1,...$ denote the unique sequence of games
of $G$ such that $R^{0}=G,$ $R^{t}\Rightarrow R^{t+1}$ is fast $\forall t$
and $H_{i}=\cap _{t}R_{i}^{t}$ for each $i\in I.$ Suppose that $x^{\ast }\in
\tprod_{i\in I}G_{i}$ is a maximal element in the game $G,$ that is $%
P_{i}(x^{\ast })=\emptyset $ and then, $x_{i}^{\ast }$ is never eliminated
in the sequence $R^{t}$ $\forall i\in I.$ It follows that $x^{\ast }\in
\tprod_{i\in I}H_{i},$ so that $P_{i|\tprod_{i\in I}H_{i}}(x^{\ast
})=\emptyset $ and, therefore, $P_{i|\tprod_{i\in I}H_{i}}(x^{\ast })\cap
H_{i}=\emptyset $ and $x$ is also a maximal element in game $H.$

Conversely, let $x^{\ast }\in \tprod_{i\in I}H_{i}$ be a maximal element in
game $H$ $(P_{i}(x^{\ast })\cap H_{i}=\emptyset $ for each $i\in I)$ and
consider $z^{\ast }$ as in the hypothesis: $z^{\ast }\in \tprod_{i\in
I}G_{i} $ such that $z_{i}^{\ast }\in Q_{i}(z,x_{-i}^{\ast })$ for all $z\in
\tprod_{i\in I}G_{i}.$ We will prove that $P_{i}(z_{i}^{\ast },x_{-i}^{\ast
})=\emptyset .$ If we assume, on the contrary, that there exists $%
x_{i}^{\prime }\in P_{i}(z_{i}^{\ast },x_{-i}^{\ast }),$ it follows,
according to property T, $Q_{i}(x_{i}^{\prime },x_{-i}^{\ast })\subset
P_{i}(z_{i}^{\ast },x_{-i}^{\ast }).$ However, $z_{i}^{\ast }\in
Q_{i}(z,x_{-i}^{\ast })$ for all $z\in \tprod_{i\in I}G_{i},$ particularly $%
z_{i}^{\ast }\in Q_{i}(x_{i}^{\prime },x_{-i}^{\ast })$and then, $%
z_{i}^{\ast }\in P_{i}(z_{i}^{\ast },x_{-i}^{\ast }),$ which contradicts the
hypothesis$.$ Since $P_{i}(z_{i}^{\ast },x_{-i}^{\ast })=\emptyset ,$ $%
z_{i}^{\ast }$ is never eliminated in the sequence $R^{t}$ $\forall i\in I,$
and $z^{\ast }\in \tprod_{i\in I}H_{i}.$ The last assertion implies $%
z_{i}^{\ast }\in Q_{i}(z_{i},x_{-i}^{\ast })\cap H_{i}$ for all $z\in
\tprod_{i\in I}G_{i}.$ We will prove that $P_{i}(x^{\ast })=\emptyset .$ Let
us assume, on the contrary, that there exists $x^{\prime }\in \tprod_{i\in
I}G_{i}$ such that $x_{i}^{\prime }\in P_{i}(x^{\ast })$ for each $i\in I.$
Then, according to Property $T,$ we have that $Q_{i}(x_{i}^{\prime
},x_{-i}^{\ast })\subset P_{i}(x^{\ast }).$ However, we have that $%
z_{i}^{\ast }\in Q_{i}(x_{i}^{\prime },x_{-i})$ from hypothesis, so that $%
z_{i}^{\ast }\in P_{i}(x^{\ast }).$ In addition, $z_{i}^{\ast }\in H_{i}$
and, then, $z_{i}^{\ast }\in P_{i}(x^{\ast })\cap H_{i},$ which contradicts
the fact that $P_{i}(x^{\ast })\cap H_{i}=\emptyset .$ In conclusion, $%
P_{i}(x^{\ast })$ must be the empty set and $x^{\ast }$ is a maximal element
for game $G.$ $\square \medskip $

For the $(\rightarrow ^{\ast })-$reductions of a game, we obtain the
following corollary.

\begin{corollary}
Let $G=(G_{i},P_{i},Q_{i})_{i\in I}$ be a general qualitative game which
satisfies property $T$ and $x_{i}\notin P_{i}(x_{i},x_{-i})$ for each $%
x_{-i}\in \tprod_{i\in I}G_{i}.$ Let us assume that for each $x\in
\tprod_{i\in I}G_{i},$ there exists $z^{\ast }\in \tprod_{i\in I}G_{i}$ such
that $z_{i}^{\ast }\in Q_{i}(z_{i},x_{-i})$ for all $z\in \tprod_{i\in
I}G_{i}$ and $i\in I.$ If $H$ is a $(\rightarrow ^{\ast })-$reduction of $G,$
then games $G$ and $H$ have the same maximal elements.
\end{corollary}

\section{CONCLUDING REMARKS}

Dufwenberg and Stegeman [8] let an open problem for researchers, that is to
find classes of games for which order independence holds. We reconsidered
the problem of the existence of nonempty maximal reductions for qualitative
games, which can be generalizations of discontinuous strategic games. We
generalized the results obtained by Dufwenberg and Stegeman [8] and Apt [1].

\end{document}